
\input harvmac

\overfullrule=0pt


\def\J{{\scriptscriptstyle J}}

\def\Q{{\scriptscriptstyle Q}}

\def\S{{\scriptscriptstyle S}}

\def\Z{{\scriptscriptstyle Z}}


\def\CA{{\cal A}}

\def\CO{{\cal O}}


\def\a{\alpha}
\def\b{\beta}
\def\c{\gamma}
\def\d{\delta}
\def\e{\epsilon}

\def\n{\eta}
\def\o{\sigma}

\def\s{\sigma}
\def\t{\tau}
\def\th{\theta}
\def\u{\mu}
\def\v{\nu}


\def\aS{\alpha_s}
\def\bar#1{\overline{#1}}
\def\bbar{{\overline b}}
\def\bbbaroctetsingS{b\bbar[^1S_0^{(8)}]}
\def\bbbaroctettripS{b\bbar[^3S_1^{(8)}]}
\def\bbbaroctettripP{b\bbar[^3P_J^{(8)}]}

\def\Br{{\rm Br}}
\def\cbar{{\overline c}}

\def\ccbaroctetsingS{c\cbar[ ^1S_0^{(8)}]}
\def\ccbaroctettripP{c\cbar[ ^3P_J^{(8)}]}

\def\ccbaroctettripS{c\cbar[ ^3S_1^{(8)}]}

\def\ccdot{\hbox{\kern-.1em$\cdot$\kern-.1em}}
\def\chibJ{\chi_{b\J}}
\def\chicJ{\chi_{c\J}}
\def\chiQJ{\chi_{\Q\J}}


\def\GeV{{\>\, \rm GeV}}

\def\gS{g_s}
\def\gtap{ \> \raise.3ex\hbox{$>$\kern-.75em\lower1ex\hbox{$\sim$}} \> }
\def\Jpsi{J/\psi}
\def\JZ{J_\Z}

\def\ltap{\raise.3ex\hbox{$<$\kern-.75em\lower1ex\hbox{$\sim$}}}
\def\LZ{L_\Z}
\def\mb{m_b}
\def\Mb{M_b}
\def\mc{m_c}
\def\Mc{M_c}

\def\MQ{M_\Q}

\def\Nc{N_c}
\def\NDOF{{\rm NDOF}}

\def\octet{{\rm octet}}

\def\pbar{{\overline{p}}}

\def\pperp{p_\perp}
\def\psiQ{\psi_\Q}
\def\pslash{p\hskip-0.5em /} 
 
\def\pT{p_\perp}
\def\qbar{{\overline q}}
\def\Qbar{{\overline Q}}

\def\qhat{{\hat{q}}}
\def\QQbar{{Q\Qbar}}
\def\QQbaroctetgen{Q\Qbar[ ^{2S+1}L_J^{(8)}]}
\def\QQbaroctetPzero{Q\Qbar[^3P_0^{(8)}]}
\def\QQbaroctetPone{Q\Qbar[^3P_1^{(8)}]}

\def\QQbaroctetPtwo{Q\Qbar[^3P_2^{(8)}]}

\def\QQbaroctetgen{Q\Qbar[ ^{2S+1}L_J^{(8)}]}
\def\QQbaroctetsingS{Q\Qbar[ ^1S_0^{(8)}]}
\def\QQbaroctettripP{Q\Qbar[ ^3P_J^{(8)}]}
\def\QQbaroctettripPone{Q\Qbar[ ^3P_1^{(8)}]}

\def\QQbaroctettripS{Q\Qbar[ ^3S_1^{(8)}]}

\def\QQbarpairgen{Q\Qbar[ ^{2S+1}L_J^{(1,8)}]}
\def\QQbarsingletS{Q\Qbar[ ^1S_0^{(1)}]}
\def\qslash{q\hskip-0.5em /} 
\def\shat{\hat{s}}
\def\short{{\rm short}}

\def\SZ{S_\Z}

\def\that{\hat{t}}

\def\ubar{\bar{u}}
\def\uhat{\hat{u}}
\def\vbar{\bar{v}}
\def\vb{v_b}
\def\vc{v_c}
\def\vQ{v_\Q}
\def\zhat{\hat{z}}


\def\half{{1 \over 2}}

\def\third{{1 \over 3}}

\def\twothirds{{2 \over 3}}


\newdimen\pmboffset
\pmboffset 0.022em
\def\oldpmb#1{\setbox0=\hbox{#1}%
 \copy0\kern-\wd0 \kern\pmboffset\raise
 1.732\pmboffset\copy0\kern-\wd0 \kern\pmboffset\box0}
\def\pmb#1{\mathchoice{\oldpmb{$\displaystyle#1$}}{\oldpmb{$\textstyle#1$}}
      {\oldpmb{$\scriptstyle#1$}}{\oldpmb{$\scriptscriptstyle#1$}}}


%
%
\def\appendix#1#2{\global\meqno=1\global\subsecno=0\xdef\secsym{\hbox{#1.}}
\bigbreak\bigskip\noindent{\bf Appendix. #2}\message{(#1. #2)}
\writetoca{Appendix {#1.} {#2}}\par\nobreak\medskip\nobreak}


\nref\Bodwin{G.T. Bodwin, E. Braaten and G.P. Lepage, Phys. Rev. {\bf D51} 
(1995) 1125.}
\nref\Lepage{G.P. Lepage, L. Magnea, C. Nakhleh, U. Magnea and K. 
 Hornbostel, Phys. Rev. {\bf D46} (1992) 4052.}
\nref\Chang{C.H. Chang, Nucl. Phys. {\bf B172} (1980) 425.}
\nref\Berger{E.L. Berger and D. Jones, Phys. Rev. {\bf D23} (1981) 1521.}
\nref\Kuhn{J. H. K\"uhn, J. Kaplan and E. G. O. Safiani, Nucl. Phys. {\bf
 B157} (1979) 125.}
\nref\Guberina{B. Guberina, J.H. K\"uhn, R.D. Peccei and R. R\"uckl, Nucl. 
 Phys. {\bf B174} (1980) 317.}
\nref\Baier{R. Baier and R. R\"uckl, Z. Phys. C {\bf 19} (1983) 251.}
\nref\ChoLeibov{P. Cho and A.K. Leibovich, Phys. Rev. {\bf D53} (1996) 150.}
\nref\Tang{W.-K. Tang and M. V\"anttinen, SLAC-PUB-95-6931 (1995), 
  unpublished.}
\nref\Psidata{The CDF collaboration, Fermilab-Conf-94/136-E (1994), 
 unpublished.}
\nref\Papadimitriou{The CDF collaboration, Fermilab-conf-95/128-E (1995), 
 unpublished.}
\nref\Upsilondata{The CDF collaboration, Fermilab-Pub-95/271-E (1995), 
 unpublished.}
\nref\Quigg{E.J. Eichten and C. Quigg, Phys. Rev. {\bf D52} (1995) 1726.}
\nref\Mertig{R. Mertig, M. B\"ohm and A. Denner, Comp. Phys. Comm. {\bf 64}
 (1991) 345.}
\nref\BraatenYuanI{E. Braaten and T.C. Yuan, Phys. Rev. Lett. {\bf 71} (1993)
 1673.}
\nref\BraatenYuanII{E. Braaten and T.C. Yuan, Phys. Rev. {\bf D50} (1994)
 3176.}
\nref\BDFM{E. Braaten, M.A. Doncheski, S. Fleming and M.L. Mangano,
 Phys. Lett. {\bf B333} (1994) 548.}
\nref\Roy{D.P. Roy and K. Sridhar, Phys. Lett. {\bf B339} (1994) 141.}
\nref\Cacciari{M. Cacciari and M. Greco, Phys. Rev. Lett. {\bf 73} (1994) 
 1586.}
\nref\BraatenFleming{E. Braaten and S. Fleming, Phys. Rev. Lett. {\bf 74} 
 (1995) 3327.}
\nref\CGMP{M. Cacciari, M. Greco, M.L. Mangano and A. Petrelli, 
 Phys. Lett. {\bf B356} (1995), 553.}
\nref\UAone{The UA1 collaboration, Phys. Lett. {\bf B256} (1991) 112.}
\nref\CKY{K. Cheung, W.-Y. Keung and T.C. Yuan, Phys. Rev. Lett. {\bf 76} 
 (1996) 877.}
\nref\Cho{P. Cho, Phys. Lett. {\bf B368} (1996) 171.} 
\nref\BraatenChen{E. Braaten and Y.-Q. Chen, Phys. Rev. Lett. {\bf 76} (1996) 
 730.}
\nref\Ko{P. Ko, J. Lee and H.S. Song, Phys. Rev. {\bf D53} (1996) 1409.}
\nref\Beneke{M. Beneke and I.Z. Rothstein, SLAC-PUB-7129 (1996), 
 unpublished.}
\nref\BraatenChenII{E. Braaten and Y.-Q. Chen, OHSTPY-HEP-T-96-010 (1996), 
 unpublished.}


\nfig\qqbarQQbargraphs{Lowest order Feynman graph which mediates 
$q + \qbar \to Q +\Qbar$ scattering.}
\nfig\ggQQbargraphs{Lowest order Feynman graphs which mediate 
$g + g \to Q +\Qbar$ scattering.}
\nfig\Twototwographs{Feynman diagrams which mediate (a) $q\qbar \to \psiQ g$, 
(b) $g q \to \psiQ q$ and (c) $gg \to \psiQ g$ scattering through intermediate 
$\QQbaroctettripP$ and $\QQbaroctetsingS$ pairs.}
\nfig\PtoSratio{Ratio $R(\pT)$ of the total $\QQbaroctettripP$ and 
$\QQbaroctetsingS$ contributions to the $\psiQ$ transverse momentum 
differential cross section in the limit where the long distance NRQCD matrix 
element $\langle \CO_8^{\psiQ}(^3P_0) \rangle$ equals 
$\MQ^2 \langle \CO_8^{\psiQ}(^1S_0) \rangle$. The solid and dotted curves 
illustrate $R(\pT)$ for the charmonia and bottomonia sectors respectively.}
\nfig\Psipxsect{Theoretical transverse momentum differential cross section for
prompt $\psi'$ production at the Tevatron in the pseudorapidity interval
$|\eta| \le 0.6$ compared against preliminary CDF data.  The dashed curve 
depicts the direct color-singlet contribution to $\psi'$ production.  The 
dot-dashed curve illustrates the $\ccbaroctettripS$ cross section, and the 
dotted curve denotes the combined $\ccbaroctettripP$ and 
$\ccbaroctetsingS$ distributions.  The solid curve equals the sum of the 
color-singlet and color-octet contributions and represents the total 
theoretical prediction for the $\psi'$ differential cross section.  All 
curves are multiplied by the muon branching fraction $\Br(\psi' \to \mu^+ 
\mu^-)$.}
\nfig\Jpsixsect{Theoretical transverse momentum differential cross
section for prompt $\Jpsi$ production at the Tevatron in the pseudorapidity
interval $|\eta| \le 0.6$ compared against preliminary CDF data.  The curves 
in this figure are labeled in the same way as those in \Psipxsect.  All
curves are multiplied by the muon branching fraction $\Br(\Jpsi \to \mu^+ 
\mu^-)$.}
\nfig\ChicJxsect{Theoretical transverse momentum differential cross section 
for $\Jpsi$ production at the Tevatron in the pseudorapidity interval 
$|\eta| \le 0.6$ resulting from radiative $\chicJ$ decay compared against 
preliminary CDF data.  The dashed curve depicts the color-singlet contribution, 
the dot-dashed curve illustrates the $\ccbaroctettripS$ cross section and the 
solid curve represents their sum.  All curves are multiplied by the muon 
branching fraction $\Br(\Jpsi \to \mu^+ \mu^-)$.}
\nfig\UpsilononeSxsect{Theoretical transverse momentum differential cross
section for $\Upsilon(1S)$ production at the Tevatron in the rapidity interval
$|y| \le 0.4$ compared against preliminary CDF data.  The dashed curve
depicts the color-singlet contribution which includes direct $\Upsilon(1S)$
production as well as radiative feeddown from $\chibJ(1P)$ and $\chibJ(2P)$
states.  The dot-dashed curve illustrates the $\bbbaroctettripS$ 
cross section, and the dotted curve denotes the combined 
$\bbbaroctettripP$ and $\bbbaroctetsingS$ distributions.  The solid 
curve equals the sum of the color-singlet and color-octet contributions and 
represents the total theoretical prediction for the $\Upsilon(1S)$ differential 
cross section.  All curves are multiplied by the muon branching fraction 
$\Br(\Upsilon(1S) \to \mu^+ \mu^-)$.}
\nfig\UpsilontwoSxsect{Theoretical transverse momentum differential cross
section for $\Upsilon(2S)$ production at the Tevatron in the rapidity 
interval $|y| \le 0.4$ compared against preliminary CDF data.  The curves in 
this figure are labeled the same as those in \UpsilononeSxsect.  The dashed
color-singlet cross section includes $\Upsilon(2S)$ production and radiative
feeddown from $\chibJ(2P)$.  All curves are multiplied by the muon branching 
fraction $\Br(\Upsilon(2S) \to \mu^+ \mu^-)$.}
%


\def\CITTitle#1#2#3{\nopagenumbers\abstractfont
\hsize=\hstitle\rightline{#1}
\vskip 0.4in\centerline{\titlefont #2} \centerline{\titlefont #3}
\abstractfont\vskip .4in\pageno=0}

\CITTitle{{\baselineskip=12pt plus 1pt minus 1pt
  \vbox{\hbox{CALT-68-2026}\hbox{DOE RESEARCH AND}\hbox{DEVELOPMENT
  REPORT}}}}
{Color-Octet Quarkonia Production II}{}
\centerline{
  Peter Cho\footnote{$^1$}{Work supported in part by a DuBridge Fellowship 
and
  by the U.S. Dept. of Energy under DOE Grant no. DE-FG03-92-ER40701.}
  and Adam K. Leibovich\footnote{$^2$}{Work supported in part by
  the U.S. Dept. of Energy under DOE Grant no. DE-FG03-92-ER40701.}}
\centerline{Lauritsen Laboratory}
\centerline{California Institute of Technology}
\centerline{Pasadena, CA  91125}

\vskip .2in
\centerline{\bf Abstract}
\bigskip\bigskip

	We calculate the lowest order hadronic cross sections for producing 
colored heavy quark-antiquark pairs in $L=S=0$ and $L=S=1$ configurations.  
Such $Q\Qbar[{}^1S_0^{(8)}]$ and $Q\Qbar[{}^3P_J^{(8)}]$ states hadronize into 
$\psi_\Q$ quarkonia at the same order in the NRQCD velocity expansion as 
previously considered $Q\Qbar[{}^3S_1^{(8)}]$ pairs.  Their contributions to 
prompt Psi and Upsilon production at the Tevatron bring the shapes of 
theoretical transverse momentum distributions into line with recent CDF 
measurements.  We find that the best fit values for the linear combinations of 
$Q\Qbar[{}^1S_0^{(8)}]$ and $Q\Qbar[{}^3P_J^{(8)}]$ long distance matrix 
elements which can be extracted from the data are generally consistent with 
NRQCD scaling rules.  

\Date{4/96}

\newsec{Introduction}

	The study of quarkonia has yielded valuable insight into the nature of 
the strong interaction ever since the discovery of the $\Jpsi$ resonance in 
1974.   During the past two decades, $Q\Qbar$ bound states have provided 
useful laboratories for probing both perturbative and nonperturbative aspects 
of QCD.  Recently, investigations of charmonia and bottomonia systems have 
uncovered some striking surprises.  Orders of magnitude disagreements have 
been found between old predictions and new measurements of Psi and Upsilon 
production at several collider facilities.  These large disparities have 
called into question the simplest model descriptions of quarkonia and 
stimulated the development of a new paradigm for treating heavy 
quark-antiquark systems based upon QCD.  Although much theoretical and 
experimental work remains to be done before a truly consistent picture 
of quarkonia is established, ongoing studies of these heavy mesons are leading 
to a better understanding of some basic aspects of strong interaction physics. 

	Quarkonia bound states are qualitatively different from most 
other hadrons since they are inherently nonrelativistic.  The physics of 
quarkonia consequently involves several energy scales which are separated by 
the small velocity $v$ of the heavy constituents inside 
$Q\Qbar$ bound states.  The most important scales are set by the mass $\MQ$, 
momentum $\MQ v$ and kinetic energy $\MQ v^2$ of the heavy quark and 
antiquark.  In order to keep track of this scale hierarchy, an effective field 
theory called Nonrelativistic Quantum Chromodynamics (NRQCD) has been 
established \Bodwin.  This effective theory for $Q\Qbar$ bound states shares 
several similarities with the Heavy Quark Effective Theory (HQET) which 
describes the low energy QCD structure of heavy-light $Q\qbar$ mesons.  For 
example, NRQCD is based upon a double power series expansion in the strong 
interaction fine structure constant $\aS = \gS^2 / 4\pi$ and the velocity 
parameter $v \sim 1/\log\MQ$ which is similar to the HQET's double expansion 
in $\aS$ and $1/\MQ$.  Both theories also incorporate approximate spin symmetry 
relations which constrain various multiplet structures and transition rates.  
But most importantly, NRQCD systematizes one's understanding of charmonia and 
bottomonia just as the HQET methodically organizes the physics of $D$ 
and $B$ mesons. 

	Quarkonia are described within the NRQCD framework in terms of 
Fock state decompositions.  The wavefunction of an S-wave orthoquarkonium 
vector meson schematically looks like
\eqn\psiwavefunc{\eqalign{
|\, \psiQ \rangle &= O(1) |\, Q\Qbar\,[{}^3S_1^{(1)}] \,\rangle
+ O(v) |\, Q\Qbar\, [ {}^3P_J^{(8)}] \,  g \rangle \cr
& \quad 
+ O(v^2) |\, Q\Qbar\, [ {}^1S_0^{(8)}] \,  g \rangle
+ O(v^2) |\, Q\Qbar\, [ {}^3S_1^{(1,8)}] \,  g g \rangle 
+ O(v^2) |\, Q\Qbar\, [ {}^3D_J^{(1,8)}] \,  g g \rangle 
+ \cdots.}}
The spin, orbital and total angular momentum quantum numbers of the $Q\Qbar$ 
pairs in each Fock component are indicated within the square brackets in 
spectroscopic notation, while the pairs' color assignments are specified
by singlet or octet superscripts.  The order in the velocity expansion 
at which each of these Fock states participates in $\psiQ$ annihilation or 
creation processes is governed by simple NRQCD counting rules \Lepage.
For instance, suppose a heavy quark and antiquark are produced almost 
on-shell with nearly parallel 3-momenta in some high energy reaction.  The 
low energy hadronization of this pair into a physical $\psiQ$ bound 
state takes place at $O(v^3)$ if it has the same angular momentum and color 
quantum numbers as those displayed in the first Fock component of 
\psiwavefunc.  The long distance evolution of all other $Q\Qbar$ pairs 
generated at short distance scales into $\psiQ$ mesons occurs at higher orders 
in the velocity expansion.  

	If the relative importance of various quarkonia production channels 
depended solely upon the order in $v$ at which pairs hadronize into physical 
bound states, those modes which proceed through the leading Fock components in 
quarkonia wavefunction decompositions would generally be dominant.  This 
assumption coincides with the basic tenet of the so-called color-singlet model 
\refs{\Chang{--}\Baier}.  Quarkonium production is presumed in this model to 
be mediated by parton reactions that generate colorless heavy quark-antiquark 
pairs with the same quantum numbers as the mesons into which they 
nonperturbatively evolve.  Transverse momentum distributions calculated within 
this picture badly underestimate experimental observations for $\pT \gtap 
2 \MQ$.  The breakdown of the color-singlet model stems from its neglect of 
all high energy processes that create $Q\Qbar$ pairs with quantum numbers 
different from those of the final state meson.  In particular, it overlooks 
short distance contributions to quarkonia cross sections from intermediate 
color-octet states which can be orders of magnitude larger than those from 
color-singlet pairs.  Even if the long distance hadronization of the former is 
suppressed by several powers of $v$ compared to the latter, the color-octet 
components of quarkonia distributions can dominate overall. 

	In a previous paper \ChoLeibov, we examined the contributions to 
hadron collider quarkonia cross sections from processes which create colored 
heavy quark-antiquark pairs in $L~=~0$, $S~=~1$ configurations at short 
distance scales.  Such $\QQbaroctettripS$ pairs may emit or absorb a soft 
chromoelectric dipole gluon and evolve at $O(v^5)$ into $\chiQJ$ mesons.  
These P-wave quarkonia can later radiatively decay to lower S-wave $\psiQ$ 
levels.  Alternatively, $\QQbaroctettripS$ pairs may undergo double 
chromoelectric dipole transitions and directly hadronize at $O(v^7)$ into 
$\psiQ$ mesons.   The first channel formally 
represents the dominant color-octet production mechanism.  But since 
$\chiQJ \to \psiQ + \gamma$ branching ratios in both the charmonia and 
bottomonia sectors are numerically comparable to $v_c^2 \simeq 0.23$ and $v_b^2 
\simeq 0.08$, the second mode is phenomenologically important as well.  
Consistency then requires that $O(v^7)$ color-octet contributions to $\psiQ$ 
production from colored $L=S=0$ and $L=S=1$ states also be considered.  
Therefore, we calculate in this article the $\psiQ$ cross sections arising from 
$\QQbaroctetsingS$ and $\QQbaroctettripP$ intermediate channels.   As we shall 
see, inclusion of a fully consistent set of color-octet quarkonia cross 
sections brings theoretical predictions into line with experimental 
observations of prompt Psi and Upsilon production at Fermilab.

	Our paper is organized as follows.  In section~2, we compute the 
amplitudes for $gg \to \QQbaroctetsingS$ and $gg \to \QQbaroctettripP$ 
scattering which mediate $\psiQ$ production at small transverse momenta.  We 
then utilize these amplitudes in section~3 to determine the $\QQbaroctetsingS$ 
and $\QQbaroctettripP$ contributions to $\psiQ$ differential cross sections 
in $2 \to 2$ collisions.  In section~4, we fit color-octet quarkonia 
distributions to recent Psi and Upsilon measurements and determine improved 
numerical values for several NRQCD matrix elements.  Finally, we summarize our 
findings in section~5 and close with some thoughts on the implications of our 
results for quarkonia production in several different experimental settings.

\newsec{$\pmb{\psiQ}$ production in $\pmb{2 \to 1}$ collisions}

	Color-octet quarkonia production starts at $O(\aS^2)$ with the 
scattering processes $q + \qbar \to Q + \Qbar$ and $g+g \to Q + \Qbar$.
The Feynman diagrams which mediate these reactions are illustrated in 
\qqbarQQbargraphs\ and \ggQQbargraphs.  In the first quark scattering 
channel, the heavy quark-antiquark pair appearing in the final state 
must share the same quantum numbers as its intermediate virtual gluon 
progenitor.  Angular momentum, parity and charge conjugation conservation 
restrict the spin, orbital and total angular momentum quantum numbers of the 
$Q\Qbar$ pair to $L=0 \> {\rm or} \> 2$, $S=1$ and $J=1$.
%
%
In the second gluon scattering process, the allowed values for $L$, $S$ and 
$J$ are not so tightly prescribed.  In ref.~\ChoLeibov, we examined the 
important special case where gluon fusion produces $\QQbaroctettripS$ pairs. 
We now generalize our earlier analysis and consider the $O(\aS^2)$ formation 
of other pairs with different sets of quantum numbers.

	To begin, we write down the on-shell scattering amplitude 
\eqn\ggQQbaramp{\eqalign{
& \CA\left( g_a(p_1) g_b(p_2) \to Q^i({P \over 2}+q;s_1) 
  \Qbar_j({P \over 2}-q;s_2) \right) =
-2 \gS^2 \varepsilon^\mu(p_1) \varepsilon^\nu(p_2) \cr
& \quad \times \ubar({P \over 2}+q;s_1) \Bigl\{ 
(T_a T_b)^i_j {\gamma_\mu [\pslash_2-\pslash_1+2 \qslash+2 \MQ] \gamma_\nu 
  \over (p_2-p_1+2q)^2 - 4 \MQ^2} 
+(T_b T_a)^i_j  {\gamma_\nu [\pslash_1-\pslash_2+2 \qslash+2 \MQ] \gamma_\mu 
  \over (p_2-p_1-2q)^2 - 4 \MQ^2} \cr
& \qquad\qquad\qquad\qquad -{i \over 8\MQ^2} f_{abc}(T_c)^i_j \bigl[ g_{\mu\nu} 
(\pslash_2-\pslash_1) + 2 (p_{1\nu} \gamma_\mu - p_{2\mu} \gamma_\nu) \bigr] 
\Bigr\} v({P \over 2}-q;s_2) \cr}}
that corresponds to the sum of the three graphs displayed in \ggQQbargraphs.
The reduced amplitude which describes the creation of a $Q\Qbar$ pair in a 
particular angular momentum and color configuration is obtained from this 
expression by applying a series of projection operations:
\eqn\projection{\eqalign{
& \CA \left( g_a(p_1) g_b(p_2) \to Q\Qbar[ ^{2S+1}L_\J^{(1,8c)}](P) \right) 
= \cr
& \quad \sum_{\LZ,\SZ} \sum_{s_1,s_2} \sum_{i,j} 
\int {d^3 q \over (2\pi)^3 2 q^0} \d\bigl( q^0 - {{\vec q}^{\, 2} \over M} 
\bigr)  Y^*_{L\LZ}(\qhat) \langle \half s_1; \half s_2 | S \SZ \rangle
\langle L \LZ ; S \SZ | J \JZ \rangle
\langle 3  i ; \bar{3} j | 1,8 c \rangle \cr
& \qquad\qquad\qquad \times 
\CA\left(g_a(p_1) g_b(p_2) \to Q^i({P \over 2}+q;s_1) 
\Qbar_j({P \over 2}-q;s_2) \right). \cr}}
Several points about this projection formula should be noted.  Firstly, a 
$Q\Qbar$ pair has negligible overlap with a nonrelativistic quarkonium bound 
state unless the relative momentum $q$ between the heavy quark and antiquark is 
small compared to their combined momentum $P$.  We have therefore incorporated 
a delta function into \projection\ which restricts the triple integral over 
$q$ to the two-dimensional surface defined by $|{\vec q}\, | = \sqrt{M q^0}$ 
where $q^0 \simeq M v^2 \ll M \equiv 2 \MQ$.  The squared invariant mass of 
the $Q\Qbar$ pair thus equals $P^2 = M^2$ up to small relativistic 
corrections.  Inclusion of the delta function also 
properly converts the mass dimension of the $2 \to 2$ scattering amplitude into 
that for a $2 \to 1$ reaction.  Secondly, the angular integration over the 
spherical harmonic projects out the pair's specified partial wave.  The sums 
over the Clebsch-Gordan coefficients similarly project out the spin 
and total angular momentum of the $Q\Qbar$ object.  Finally, the sum 
over the $SU(3)$ coefficients 
\eqn\colorcoeffs{\eqalign{
\langle 3  i ; \bar{3} j | 1 \rangle &= {\d^j_i / \sqrt{\Nc}} \cr
\langle 3  i ; \bar{3} j | 8 c \rangle &= \sqrt{2} (T_c)^j_i \cr}}
combines together the color quantum numbers of the quark and antiquark into 
either a singlet or octet configuration. 

	Inserting the gluon scattering amplitude \ggQQbaramp\ into 
the projection formula in \projection, we can readily calculate the 
reduced amplitude for two gluons to fuse into an arbitrary color-octet 
combination.  We list below the formation amplitudes for $\QQbaroctettripS$, 
$\QQbaroctetsingS$ and $\QQbaroctettripP$ pairs which all hadronize into 
$\psiQ$ bound states at $O(v^7)$ in the velocity expansion:
\eqna\ggQQbaramps
$$ \eqalignno{
& \CA(g_a(p_1) g_b(p_2) \to \QQbaroctettripS_c) = {i \over 4 (2\pi)^3} 
\sqrt{4 \pi M \over q^0}  g_s^2 
 f_{abc} \varepsilon^\mu(p_1) \varepsilon^\nu(p_2) 
 \varepsilon^\sigma(S_z)^* \cr
& \qquad \times {p_1^2 + p_2^2 \over p_1^2 + p_2^2 -M^2}  
\bigl[ (p_2-p_1)_\sigma g_{\u\v} + 2 (p_{1\v} g_{\u\o} - p_{2\u} g_{\v\o})
\bigr] & \ggQQbaramps a \cr 
& \CA(g_a(p_1) g_b(p_2) \to \QQbaroctetsingS_c) = 
- {i \over 2 (2\pi)^3} \sqrt{4 \pi M \over q^0} \gS^2  d_{abc} 
{M \epsilon_{\s\t\mu\nu} p_1^\s p_2^\t \varepsilon^\mu(p_1) 
\varepsilon^\nu(p_2) \over p_1^2 + p_2^2 - M^2}
& \ggQQbaramps b \cr
& \CA(g_a(p_1) g_b(p_2) \to \QQbaroctettripP_c) = {1 \over 2 (2\pi)^3} 
\sqrt{4 \pi \over 3} \gS^2 d_{abc} \sum_{\LZ,\SZ} 
\langle 1 \LZ ; 1 \SZ | J \JZ \rangle 
\varepsilon^\a(\LZ)^* \varepsilon^\b(\SZ)^* \cr
& \qquad \times {M \varepsilon^\mu(p_1) \varepsilon^\nu(p_2) \over
p_1^2 + p_2^2 - M^2} \Bigl\{ g_{\a\u} p_{1\v} (p_2-p_1)_\b + g_{\a\v} p_{2\u} 
(p_1-p_2)_\b \cr
& \qquad\qquad\qquad\qquad\qquad + (p_2^2-p_1^2-M^2) g_{\a\v} g_{\b\u} 
 + (p_1^2-p_2^2-M^2) g_{\a\u} g_{\b\v} \cr
& \qquad\qquad\qquad\qquad\qquad + {M^2 \over M^2-p_1^2-p_2^2} 
 \bigl[ g_{\u\v} p_{1\b} (p_1-p_2)_\a + g_{\u\v} p_{2\b} (p_2-p_1)_\a \bigr] \cr
& \qquad\qquad\qquad\qquad\qquad +{M^2+p_1^2 + p_2^2 \over M^2 - p_1^2 - p_2^2}
\bigl[ g_{\u\b} p_{1 \v} (p_2-p_1)_\a + g_{\v\b} p_{2\u} (p_1-p_2)_\a \bigr]
\Bigr\}. & \ggQQbaramps c \cr} $$
As required by gauge invariance, these expressions vanish when 
$p_1^2 = p_2^2=0$ and $\varepsilon^\u(p_1) \to p_1^\u$ or 
$\varepsilon^\v(p_2) \to p_2^\v$.  The general P-wave result listed in 
\ggQQbaramps{c}\ may be further reduced by employing the Clebsch-Gordan 
identities \Guberina
\eqn\Clebschsums{\eqalign{
\sum_{\LZ,\SZ} \langle 1 \LZ ; 1 \SZ | 0 0 \rangle \varepsilon^\a(\LZ)^*
 \varepsilon^\b(\SZ)^* &= \sqrt{\third} \bigl( g^{\a\b} - {P^\a P^\b \over M^2}
 \bigr) \cr
\sum_{\LZ,\SZ} \langle 1 \LZ ; 1 \SZ | 1 \JZ \rangle \varepsilon^\a(\LZ)^*
 \varepsilon^\b(\SZ)^* &= - {i \over \sqrt{2}M} \e^{\a\b\c\d} P_\c 
\varepsilon_\d(\JZ)^* \cr
\sum_{\LZ,\SZ} \langle 1 \LZ ; 1 \SZ | 2 \JZ \rangle \varepsilon^\a(\LZ)^*
 \varepsilon^\b(\SZ)^* &= \varepsilon^{\a\b}(\JZ)^*. \cr}}
We then find that $gg \to \QQbaroctettripPone$ as well as 
$gg \to \QQbaroctettripS$ scattering vanishes when both incident gluons go 
on-shell \refs{\ChoLeibov,\Tang}.

	The projection formula in \projection\ can obviously be generalized to 
other parton channels besides $gg \to \QQbarpairgen$.  We may insert 
any QCD amplitude which has a heavy quark and antiquark appearing in the final 
state and project out a reduced color-singlet or color-octet expression.
Applying this general technique to the $q\qbar \to Q\Qbar$ scattering 
process pictured in \qqbarQQbargraphs, we find 
\eqn\qqbarQQbaramp{\CA(q(p_1) \qbar(p_2) \to \QQbaroctettripS_a) = 
{1 \over 4 (2\pi)^3} \sqrt{4 \pi M \over q^0} g_s^2 \vbar(p_2) \gamma_\s T_a 
u(p_1) \varepsilon^\s(S_z)^*.}
In ref.~\ChoLeibov, we performed a less general projection operation which 
did not explicitly involve a relative momentum integration nor a partial wave 
decomposition of $\CA(q\qbar \to Q\Qbar)$.  The corresponding reduced 
amplitude listed in eqn.~(2.10) of ref.~\ChoLeibov\ thus differs from the 
result displayed above by an overall multiplicative factor.  

	The squares of $2 \to 1$ amplitudes enter into the differential 
cross section for heavy pair production
\eqn\twotoonexsect{d \sigma\left( a(p_1) b(p_2) \to \QQbarpairgen(P) 
\right) = {1 \over 4 p_1 \ccdot p_2} \mathop{{\bar{\sum}}} \left | \CA(a b \to 
\QQbarpairgen)\right|^2 d \Phi_1(p_1+p_2; P)}
where the barred summation symbol indicates that initial (final) spins and 
colors are averaged (summed) and $d \Phi_1$ denotes a one-body 
phase space factor.   High and low energy effects are intertwined 
in this expression.  In order to disentangle it, we follow ref.~\Bodwin\ and 
match the integrated cross section onto the product of a short distance 
coefficient and a long distance NRQCD matrix element: 
\eqn\matchI{\sigma\left( a b \to \QQbarpairgen \right) = 
{C_\short \over \MQ^{d-4}} \times \langle 0 | \CO_{1,8}^\QQbar(^{2S+1}L_J) | 0 
\rangle.}
The general structure of the operator whose vacuum-to-vacuum matrix element 
appears on the RHS of this matching condition looks like
\eqn\QQbarpairop{\CO_{1,8}^\QQbar(^{2S+1}L_J) = \chi^\dagger K \psi 
\left( \sum_{m_\J} \Bigl| \QQbarpairgen \Bigr\rangle \> \Bigl\langle 
\QQbarpairgen \Bigr| \right) \psi^\dagger K \chi} 
where $\psi$ and $\chi$ represent two-component Pauli spinor fields and the 
matrix $K$ denotes a product of color, spin and covariant derivative factors.
The intermediate quark-antiquark state sandwiched in the middle 
\eqn\QQbarpair{\eqalign{
& \Bigl|\QQbarpairgen \Bigr\rangle = \sum_{\LZ,\SZ} \sum_{s_1,s_2} \sum_{i,j} 
\int {d^3 q \over (2\pi)^3 2 q^0} \d\bigl( q^0 - {{\vec q}^2 \over M} \bigr) 
\cr
& \qquad\qquad \times Y^*_{L\LZ}(\qhat) 
\langle \half s_1; \half s_2 | S \SZ \rangle
\langle L \LZ ; S \SZ | J \JZ \rangle
\langle 3  i ; \bar{3} j | 1,8 \rangle \> 
\Bigl| Q^i(q; s_1) \Qbar_j(-q; s_2) \Bigr\rangle }}
is defined in the NRQCD effective theory in the same way as in full QCD. 
As a result, any arbitrariness in the definitions of the heavy pair 
production cross section and NRQCD matrix element cancels out of their ratio.  
The short distance coefficient appearing on the RHS of \matchI\ is convention 
independent. 

	All information related to the hard scattering process which creates 
the $Q\Qbar$ pair is encoded within $C_\short$.  This same coefficient enters 
into the physical quarkonium production cross section
\eqn\matchII{\sigma\left( a b \to \QQbarpairgen \to \psiQ+X\right) = 
{C_\short \over \MQ^{(d+1)-4}} \times \langle 0 | \CO_{1,8}^{\psiQ}(^{2S+1}L_J) 
| 0 \rangle.}
On the other hand, the accompanying long distance matrix element which 
specifies the probability that a $\QQbarpairgen$ pair hadronizes into a 
$\psiQ$ bound state is completely different from its counterpart in \matchI.  
The operator 
\eqn\psiQop{\CO_{1,8}^{\psiQ}(^{2S+1}L_J) = \chi^\dagger K \psi 
\left( \sum_{m_\J} \sum_X | \psiQ+X \rangle \> \langle \psiQ+X | \right) 
\psi^\dagger K \chi} 
has one unit greater mass dimension than $\CO_{1,8}^\QQbar(^{2S+1}L_J)$ as 
can be verified by comparing the dimensions of heavy intermediate pair and 
nonrelativistically normalized $\psiQ$ states.  The inverse powers of $\MQ$ in 
cross section equations \matchI\ and \matchII\ consequently differ by unity.  
The nonperturbative matrix element
$\langle 0 | \CO_{1,8}^{\psiQ}(^{2S+1}L_J) | 0 \rangle$ also cannot 
readily be calculated within NRQCD unlike its perturbative
$\langle 0 | \CO_{1,8}^{\QQbar}(^{2S+1}L_J) | 0 \rangle$ counterpart.  Simple
multiplicity relations such as 
\eqn\multiplicityrelations{\eqalign{\langle 0 | \CO^H_{1,8}(^3S_1) | 0 
\rangle &= 3 \> \langle 0 | \CO^H_{1,8}(^1S_0) | 0 \rangle\cr
\langle 0 | \CO^H_{1,8}(^3P_J) | 0 \rangle 
&= (2J+1) \langle 0 | \CO^H_{1,8}(^3P_0) | 0 \rangle\cr}}
are obeyed exactly by the latter and approximately by the former.  But the 
color-factor relation 
\eqn\colorreln{\langle 0 | \CO_8^{H}(^{2S+1}L_J) | 0 \rangle = 
{\Nc^2-1\over2\Nc} \langle 0 | \CO_1^{H}(^{2S+1}L_J) | 0 \rangle }
which holds for $H=\QQbar$ certainly does not apply when $H=\psiQ$. 
Numerical values for $\langle 0 | \CO_{1,8}^{\psiQ}(^{2S+1}L_J) | 0 \rangle$ 
matrix elements must be extracted either from experimental data or 
lattice calculations.

	In order to clarify the meaning of these NRQCD matching ideas, we 
explicitly evaluate the matching conditions specified in eqns.~\matchI\ and 
\matchII\ for one simple example.   We consider the gluon fusion formation of 
an $\eta_\Q$ pseudoscalar meson through an intermediate $\QQbarsingletS$ 
pair.  A straightforward computation yields the color-singlet cross section 
\eqn\etaxsect{\s \left( gg \to \QQbarsingletS \right) = {\aS^2 \over 384 \pi^2} 
{M \over q^0 \shat} \d \left( 1-{M^2 \over \shat} \right)}
and matrix element 
\eqn\oneSzeroME{\langle 0 | \CO_1^\QQbar({}^1S_0) | 0 \rangle = 
\Bigl\langle 0 \Bigl| \chi^\dagger \psi \Bigl| \QQbarsingletS \Bigr\rangle 
\> \Bigl\langle \QQbarsingletS \Bigr| \psi^\dagger \chi \Bigr| 0 \Bigr\rangle 
= {\Nc \over 128 \pi^5} {M^3 \over q^0}.}
We derived this last result within the NRQCD effective theory by decomposing 
the Pauli fields
\eqn\paulidecomp{\eqalign{\psi^i_\a(x) &= \sum_{s=1}^2\int{d^3 p
\over(2\pi)^3}\, b^i(p;s) \,\xi_\a(p,s)\,e^{-ip\cdot x} \cr
\chi_{i\a}(x) &= \sum_{s=1}^2\int{d^3 p
\over(2\pi)^3}\,c_i^\dagger(p;s) \,\n_\a(p,s)\,e^{ip\cdot x} \cr}}
in terms of two-component spinors normalized according 
to 
\eqn\spinors{\sum_{s=1}^2 \xi_\a(p,s)\,\xi^\dagger_\b(p,s) =
\sum_{s=1}^2 \n_\a(p,s)\,\n^\dagger_\b(p,s) = \d_{\a\b}}
and single fermion creation and annihilation operators which satisfy 
the nonrelativistic anticommutation relations
\eqn\anticommrelns{\bigl\{b^i(p;s) ,b_j^\dagger(p';s') \bigr\} = 
\bigl\{c^i(p;s) ,c_j^\dagger(p';s') \bigr\} = 
(2\pi)^3 \d^i_j \,\d_{ss'}\,\d^{(3)}(p - p').}
Taking the ratio of \etaxsect\ and \oneSzeroME, we deduce the short distance 
coefficient in matching condition \matchI
\eqn\oneSzerocoeff{ {C(gg \to \QQbarsingletS)_\short \over \MQ^{d-4}} 
= {1 \over 9} {\pi^3 \aS^2 \over M^2 \shat} \d\left(1 - {M^2\over\shat} 
\right)}
and the gluon fusion cross section in matching condition \matchII:
\eqn\newetaxsect{\s\left(gg \to \QQbarsingletS \to \eta_\Q \right) = 
{2 \over 9} {\pi^3 \aS^2 \over M^3 \shat} \d\left(1 - {M^2 \over \shat} \right) 
\langle 0 | O_1^{\eta_\Q} ({}^1S_0) | 0 \rangle.}
If we recall the relation between the NRQCD matrix element and squared 
$\eta_\Q$ wavefunction at the origin \Bodwin
\eqn\wavefuncreln{\langle 0 | O_1^{\eta_\Q} ({}^1S_0) | 0 \rangle = 
{\Nc \over 2\pi} R(0)^2 \bigl(1+O(v^4)\bigr) ,}
we see that our result is consistent with the $O(\aS^2)$ cross section 
\eqn\BRetaxsect{\s\left(gg \to \QQbarsingletS \to \eta_\Q \right) = 
{1 \over 3} {\pi^2\aS^2 \over M^3 \shat} \d\left(1 - {M^2 \over \shat} \right) 
R(0)^2}
previously reported in the literature \Baier.  

	Working in a similar fashion, we can decompose any color-singlet or 
color-octet cross section into products of short and long distance factors.  We 
tabulate in the Appendix all $O(\aS^2)$ short distance squared amplitudes for 
$2 \to 1$ color-octet reactions which yield $\psiQ$ bound states 
at $O(v^7)$ in the NRQCD velocity expansion.  The corresponding long 
distance factors are simply given by appropriate NRQCD matrix elements for 
specific production channels.  For example, the total squared amplitude for 
$gg \to \QQbaroctetsingS \to \psiQ$ scattering equals the product of the 
process-independent high energy expression listed in eqn.~(A.2a) and the 
process-specific low energy matrix element 
$\langle 0 | O_8^{\psi_\Q} ({}^1S_0) | 0 \rangle$:
\eqn\sqampexample{\mathop{{\bar{\sum}}} \left | \CA(gg \to \QQbaroctetsingS \to 
\psiQ) \right |^2 = {5 (4\pi \aS)^2 \over 192 M} 
\langle 0 | O_8^{\psi_\Q} ({}^1S_0) | 0 \rangle.}

	Color-octet pair production in $2 \to 1$ collisions could represent an 
important source of quarkonia in fixed target experiments, and its impact 
needs to be studied.  But before definite predictions can be made, numerical 
values for color-octet matrix elements must be known.  Therefore, we now turn 
to consider quarkonia production at hadron colliders where we can use 
experimental data to determine these matrix element values.

\newsec{$\pmb{\psiQ}$ production in $\pmb{2 \to 2}$ collisions}

	In order to be experimentally detectable, quarkonia must be created 
at collider facilities with nonvanishing transverse momenta so that they 
are not lost down the beampipe.  Hadrons resulting from $2 \to 1$ 
scattering processes typically have small $\pT$ comparable to the QCD scale.  
The production of quarkonia with nonnegligible transverse momenta therefore 
mainly proceeds through $2 \to 2$ collisions.   Such reactions start at 
$O(\aS^3)$ via the parton channels $q\qbar \to \QQbarpairgen g$, 
$g q \to \QQbarpairgen q$ and $g g \to \QQbarpairgen g$.  In ref.~\ChoLeibov, 
we calculated the differential production cross sections for $L=0$, $S=1$ 
color-octet pairs.  In this section, we generalize our previous results and 
consider the formation of colored $L=S=0$ and $L=S=1$ pairs. 

	The Feynman diagrams which mediate quarkonia production in these 
color-octet channels are illustrated in \Twototwographs.  The shaded circles 
appearing in the figure represent the $gg \to \QQbaroctetsingS$ and 
$gg \to \QQbaroctettripP$ amplitudes in eqns.~\ggQQbaramps{b}\ and 
\ggQQbaramps{c}.  The $q\qbar \to \psiQ g$ and $g q \to \psiQ q$ diagrams pictured in figs.~3a 
and 3b can readily be squared using standard spinor summation techniques. 
On the other hand, conventional evaluation of the gluon channel graphs 
in fig.~3c represents a formidable computational task.  It is therefore 
advisable to find a more tractable method for calculating the color-octet 
contributions to $g g \to \psiQ g$ scattering.

	We adopt a simple helicity amplitude technique to sum and square the 
gluon graphs in fig.~3c.  We first choose the following explicit 
representations for the gluon momenta and polarization vectors shown in the 
figure:
\eqn\gluonvectors{\eqalign{
p_1 &= {\sqrt{\shat} \over 2} \bigl( 1,0,0,1 \bigr) \cr
p_2 &= {\sqrt{\shat}\over2}\bigl(1,0,0,-1\bigr) \cr
p_4 &= {\shat - M^2 \over 2 \sqrt{\shat}}\bigl(1,0, \sin \th,-\cos \th \bigr)
\cr}
\qquad\qquad\eqalign{
\varepsilon_1^+ &= \varepsilon_2^- = -{1\over\sqrt{2}}(0,1,i,0)\cr
\varepsilon_1^- &= \varepsilon_2^+ = {1\over\sqrt{2}}(0,1,-i,0)\cr
(\varepsilon_4^\pm)^* &= {1\over\sqrt{2}}(0,\pm 1,i\cos\th,i\sin\th). \cr
}}
We next boost the the heavy pair's four-momentum from its primed rest frame 
to the unprimed lab frame:
\eqn\QQbarmom{p'_3 = (M,0,0,0) \to p_3 = \left( {\shat+M^2 \over 2 
\sqrt{\shat}}, 0, - {\shat-M^2 \over 2 \sqrt{\shat}} \sin\th,
{\shat-M^2 \over 2 \sqrt{\shat}} \cos\th \right).}
We also Lorentz transform the rest frame polarization vectors and tensors 
\eqn\QQbarpols{
\eqalign{(\varepsilon'_3\,^{(h=1)})^*  &=
-\sqrt{\half}\pmatrix{0\cr
		      1\cr
		      -i\cr
		      0\cr}
\cr\cr
(\varepsilon'_3\,^{(h=0)})^*  &=
\pmatrix{0\cr
	 0\cr
	 0\cr
	 1\cr}
\cr\cr
(\varepsilon'_3\,^{(h=-1)})^*  &=
\sqrt{\half}\pmatrix{0\cr
		     1\cr
		     i\cr
		     0\cr} }
\qquad\qquad\eqalign{(\varepsilon'_3\,^{(h=0)})^*  &=
\sqrt{\twothirds}\pmatrix{0&0&0&0\cr
			     0&-\half&0&0\cr
		 	     0&0&-\half&0\cr
			     0&0&0&1\cr}
 \cr\cr
(\varepsilon'_3\,^{(h=\pm1)})^*  &=
\mp\half\pmatrix{0&0&0&0\cr
		    0&0&0&1\cr
		    0&0&0&\mp i\cr
		    0&1&\mp i&0\cr}
\cr\cr
(\varepsilon'_3\,^{(h=\pm2)})^* &=
\half\pmatrix{0&0&0&0\cr
		 0&1&\mp i&0\cr
		 0&\mp i&-1&0\cr
		 0&0&0&0\cr} \cr}}
of $J=1$ and $J=2$ $\QQbar$ pairs.  Given these explicit representations, it 
is easy to work out all possible scalar contractions and express the answers 
in terms of the Mandelstam invariants $\shat$, 
$\that = -(\shat-M^2)(1-\cos\th)/2$ and 
$\uhat = -(\shat-M^2)(1+\cos\th)/2$.  The gluon channel amplitudes are 
functions of these Lorentz invariant dot products.

	Using the high energy physics package FEYNCALC \Mertig, we calculated 
each individual helicity amplitude for $gg \to \QQbaroctetsingS g$ and 
$gg \to \QQbaroctettripP g$ scattering.  Parity and crossing symmetry relations 
between different helicity amplitudes provided valuable checks on our 
Mathematica code.  Since separate helicity amplitudes do not interfere, the 
total squared amplitude simply equals the sum of the squared helicity 
amplitudes.  The final results for the $gg \to \QQbaroctetgen g$ channel are 
displayed in the Appendix alongside those for the $q\qbar \to \QQbaroctetgen g$ 
and $g q \to \QQbaroctetgen q$ modes.  For completeness, we include in this 
list the $\QQbaroctettripS$ squared amplitudes which we calculated in 
ref.~\ChoLeibov.

	The products of short distance color-octet squared amplitudes and 
long distance NRQCD matrix elements enter into the partonic cross section
\eqn\partonxsect{\eqalign{
{d\sigma\over d\that} (ab \to \QQbaroctetgen c & \to \psiQ)_{\rm octet} = \cr
& {1 \over 16 \pi \shat^2} \mathop{{\bar{\sum}}}
\left | \CA \bigl(ab \to \QQbaroctetgen c \bigr)_\short \right |^2 
\langle 0 | O_8^{\psi_\Q} ({}^{2S+1}L_J) | 0 \rangle. \cr}}
After folding in distribution functions $f_{a/A}(x_a)$ and $f_{b/B}(x_b)$ 
that specify the probabilities of finding partons $a$ and $b$ inside hadrons
$A$ and $B$ carrying momentum fractions $x_a$ and $x_b$, we obtain the 
hadronic cross section
\eqn\hadronxsect{ {d^3 \sigma \over dy_3 dy_4 d\pT}(AB \to \psiQ X)_{\rm
octet}= 2\pT \sum_{abc} x_a x_b f_{a/A}(x_a) f_{b/B}(x_b) 
{d\sigma\over d\that} (ab \to \QQbaroctetgen c \to \psiQ)_{\rm octet} }
which is a function of the $\psiQ$ and recoiling jet rapidities $y_3$ and $y_4$ 
and their common transverse momentum $\pT$.  With this hadronic 
distribution in hand, we can determine color-octet contributions to $\psiQ$ 
production in any hadronic process.  We apply it to the study of 
charmonia and bottomonia at Fermilab in the following section.

\newsec{Psi and Upsilon production at the Tevatron}

	During the past few years, striking disparities have arisen between 
old predictions and new measurements of $\Jpsi$, $\psi'$ and $\Upsilon$ 
production at the Tevatron.  The CDF collaboration has detected these heavy 
mesons at rates which exceed theoretical 
expectations based upon the color-singlet model by orders of magnitude 
\refs{\Psidata{--}\Upsilondata}.  In ref.~\ChoLeibov, we examined the 
impact of $\ccbaroctettripS$ and $\bbbaroctettripS$ intermediate states upon 
Psi and Upsilon production.  Since numerical values for most 
NRQCD color-octet matrix elements were unknown, we simply fitted the 
magnitudes of $d\s/d\pT(p\pbar \to \psiQ+X)_\octet$ cross sections to the CDF 
data.  We found that including the $\QQbaroctettripS$ channel significantly 
diminished discrepancies between the shapes of the predicted and measured 
transverse momentum distributions.  We now update our earlier analysis and 
incorporate cross section contributions from $\QQbaroctetsingS$ and 
$\QQbaroctettripP$ pairs.  As we shall see, the fully consistent $O(v^7)$ set 
of color-octet differential cross sections yields substantially improved fits 
to the data.

	We first plot in \PtoSratio\ the ratio 
\eqn\ratio{R(\pperp) =
{\displaystyle \sum_{J=0}^2\displaystyle{d\s\over d\pperp}
 \left(p\pbar \to Q\Qbar\bigl[^3P_J^{(8)}\bigr]+X \to \psiQ+X\right) \over
 \displaystyle{d\s\over d\pperp}
 \left(p\pbar \to Q\Qbar\bigl[^1S_0^{(8)}\bigr]+X \to \psiQ+X\right) }}
%
where we temporarily set $\langle\CO_8^{\psiQ}(^3P_0)\rangle = 
\MQ^2\langle\CO_8^{\psiQ}(^1S_0)\rangle$ for comparison purposes. 
\foot{The differential cross sections which enter into results displayed in 
\PtoSratio\ and all subsequent figures were calculated using the MRSD0 parton 
distribution functions evaluated at the renormalization scale $\mu = 
\sqrt{\pT^2 + M^2}$. }
The solid curve's nearly constant value $R(\pperp) \simeq 3$ for $\pperp \gtap 
5 \GeV$ indicates that the shapes of the $\ccbaroctetsingS$ and 
$\ccbaroctettripP$ differential cross sections are practically identical in 
the charmonia sector.  As a result, all fits for the NRQCD matrix elements in 
these color-octet channels become degenerate when performed over the transverse 
momentum range $5 \GeV \le \pperp \le 20 \GeV$ where $\Jpsi$ and $\psi'$ 
differential cross sections have been measured.  We consequently can only 
extract the linear combination $\langle\CO_8^{\psi}(^3P_0)\rangle/\Mc^2
+\langle\CO_8^{\psi}(^1S_0)\rangle/3$ along with 
$\langle\CO_8^{\psi}(^3S_1)\rangle$ from the CDF data.  In the bottomonia 
sector, the shapes of the $b\bbar[{}^1S_0^{(8)}]$ and $b\bbar[{}^3P_J^{(8)}]$ 
distributions are not exactly the same throughout the $0 \le \pperp \le 15 
\GeV$ interval where Upsilon data exists.  As indicated by the dot-dashed 
curve in \PtoSratio, $R(\pperp)$ varies around 5 over this transverse momentum 
range.  Yet the differences in shape between the $\bbbaroctettripP$ and 
$\bbbaroctetsingS$ contributions to the total Upsilon differential cross 
section are not sufficiently great so that a full three-parameter color-octet 
matrix element fit can be reliably performed.  So we will simply determine 
estimates for the linear combination 
$\langle\CO_8^{\Upsilon}(^3P_0)\rangle/\Mb^2 + 
\langle\CO_8^{\Upsilon}(^1S_0)\rangle/5$ along with 
$\langle\CO_8^{\Upsilon}(^3S_1)\rangle$.

	Our new fits to prompt charmonia production at the Tevatron within the 
pseudorapidity interval $|\eta| \le 0.6$ are illustrated in \Psipxsect\ and 
\Jpsixsect.  All contributions from $B$ meson decay have been removed from the 
data sets displayed in these figures, and radiative $\chicJ$ decay feeddown to 
the $\Jpsi$ differential cross section has been separated out as well.  
The dashed curves depict the direct color-singlet production predictions 
based upon the charm quark mass value $\Mc=1.48 \GeV$ and the Buchm\"uller-Tye 
charmonium wave functions at the origin tabulated in ref.~\Quigg. 
The dot-dashed and dotted curves illustrate the best fits for 
the $\ccbaroctettripS$ and combined $\ccbaroctettripP$ plus $\ccbaroctetsingS$ 
channels.  The solid curves show the sums of the color-singlet and color-octet 
components and represent the total predicted differential cross sections.  

	Following the interpolation procedure described in ref.~\ChoLeibov, 
we have included leading log corrections into the $\ccbaroctettripS$ 
differential cross sections so that they approach Altarelli-Parisi improved 
gluon fragmentation distributions for $\pT \gg \Mc$.  In the large transverse 
momentum limit, gluon fragmentation represents the dominant source of 
prompt charmonia \refs{\BraatenYuanI{--}\BraatenFleming}.  This asymptotic 
behavior can be seen in the dotdashed $\ccbaroctettripS$ curves of figs.~5 and 
6.  But throughout the $0 \le \pperp \le 20 \GeV$ region, they are not 
overwhelmingly larger than the combined $\ccbaroctettripP$ and 
$\ccbaroctetsingS$ components whose contributions to prompt charmonia 
production are sizable.  Inclusion of the latter color-octet channels into the 
total differential cross sections yields theoretical $\psi'$ and $\Jpsi$ 
distributions which fit the data quite well.  Their respective 
$\chi^2/\NDOF = 0.5$ and $\chi^2/\NDOF = 0.9$ figures-of-merit are nice and 
small.

	In \ChicJxsect, we plot the transverse momentum distribution of 
$\Jpsi$ mesons which result from radiative $\chicJ$ decay.  The dashed curve 
in the figure shows the color-singlet $\chicJ$ differential cross section 
multiplied by $\Br(\chicJ \to \Jpsi + \gamma)$ and summed over $J=0$, 1 and 
2.  The dot-dashed curve illustrates the $\ccbaroctettripS$ channel 
contribution.  The solid curve corresponds to their sum and represents the 
total $O(v^5)$ cross section prediction.  As indicated by its 
poor $\chi^2/\NDOF = 2.3$ value, this solid line does not fit the data 
well.  We believe that a better match could be achieved if subleading 
color-octet contributions were included.  The first subdominant corrections 
enter at $O(v^9)$ in the NRQCD velocity expansion from the long distance 
evolution of $\QQbaroctettripP$, $Q\Qbar[^3D_J^{(8)}]$ and 
$Q\Qbar[^1P_1^{(8)}]$ pairs into $\chiQJ$ bound states.  Since short 
distance production cross sections for the latter two pairs have not yet been 
calculated, we cannot legitimately include into \ChicJxsect\ subleading 
contributions from the first pair which we have computed. 

	We turn now to the bottomonium sector and consider Upsilon production 
at the Tevatron within the rapidity interval $|y| \le 0.4$.  Our new fits to 
CDF $\Upsilon(1S)$ and $\Upsilon(2S)$ data are displayed in 
\UpsilononeSxsect\ and \UpsilontwoSxsect.  No separation between prompt and 
delayed Upsilon sources has been experimentally performed.  The dashed 
curves in the figures therefore include both direct $\Upsilon$ production 
and radiative feeddown from $\chibJ$ states.  These color-singlet 
distributions are based upon the bottom quark mass value $\Mb = 4.88 
\GeV$ and the Buchm\"uller-Tye bottomonia wavefunctions at the origin 
tabulated in ref~\Quigg.  The dot-dashed and dotted curves illustrate the 
$\bbbaroctettripS$ and combined $\bbbaroctettripP$ plus $\bbbaroctetsingS$ 
fits.  The solid curves equal the sums of the color-singlet and color-octet 
contributions and represent the total 
$\Upsilon$ differential cross sections.  As we previously discussed in 
ref.~\ChoLeibov, the color-singlet and color-octet distributions are 
corrupted at very small transverse momenta by collinear divergences which 
should be factored into incident parton distribution functions.  Soft gluon 
effects also need to be resummed before the cross section turnover which is 
evident in \UpsilononeSxsect\ can be properly described.  Since we have not 
incorporated these effects, our cross section predictions are not trustworthy 
at low $\pT$.  We therefore exclude points in \UpsilononeSxsect\ and 
\UpsilontwoSxsect\ with $\pT \le 3.5 \GeV$ from our fits.  We then find 
$\chi^2/\NDOF = 0.3$ and $\chi^2/\NDOF = 0.9$ for the remaining points in 
these figures. 

	NRQCD power counting rules provide useful consistency checks on all
our fits.  We list in tables~I and II the numerical values for color-octet 
matrix elements which we extracted 
\nobreak{
$$ \vbox{\offinterlineskip
\def\tablerule{\noalign{\hrule}}
\hrule
\halign {\vrule#& \strut#&
\ \hfil#\hfil & \vrule#&
\ \hfil#\hfil & \vrule#&
\ \hfil#\hfil & \vrule# \cr
\tablerule%
height10pt && \omit && \omit && \omit &\cr
&& Color-Octet && Numerical && NRQCD &\cr
&& Matrix Element && Value $\left({\rm GeV}^3\right)$
&&\quad Scaling Order\quad&\cr
height10pt && \omit && \omit && \omit & \cr
\tablerule
height10pt && \omit && \omit && \omit &\cr
&& \quad $\langle 0|O_8^{J/\psi}({}^3S_1)|0\rangle$ \quad && 
\quad $(6.6 \pm 2.1)\times10^{-3} $\quad &&
\quad $\Mc^3\vc^7$ \quad &\cr
height10pt && \omit && \omit && \omit &\cr
&& \quad $\langle 0|O_8^{\chi_{c1}}({}^3S_1)|0\rangle$ \quad && 
\quad $(9.8 \pm 1.3)\times10^{-3}$\quad &&
\quad $\Mc^3\vc^5$ \quad &\cr
height10pt && \omit && \omit && \omit &\cr
&& \quad $\langle 0|O_8^{\psi'}({}^3S_1)|0\rangle$ \quad && 
\quad $(4.6 \pm 1.0)\times10^{-3}$ \quad &&
\quad $\Mc^3\vc^7$ \quad &\cr
height10pt && \omit && \omit && \omit &\cr
&& \quad $\langle 0|O_8^{\Upsilon(1S)}({}^3S_1)|0\rangle$ \quad && 
\quad $(5.9 \pm 1.9)\times10^{-3} $\quad &&
\quad $\Mb^3\vb^7$ \quad &\cr
height10pt && \omit && \omit && \omit &\cr
&& \quad $\langle 0|O_8^{\chi_{b1}(1P)}({}^3S_1)|0\rangle$ \quad && 
\quad $(4.2 \pm 1.3)\times10^{-1}$\quad &&
\quad $\Mb^3\vb^5$ \quad &\cr
height10pt && \omit && \omit && \omit &\cr
&& \quad $\langle 0|O_8^{\Upsilon(2S)}({}^3S_1)|0\rangle$ \quad && 
\quad $(4.1 \pm 0.9)\times10^{-3}$ \quad &&
\quad $\Mb^3\vb^7$ \quad &\cr
height10pt && \omit && \omit && \omit &\cr
&& \quad $\langle 0|O_8^{\chi_{b1}(2P)}({}^3S_1)|0\rangle$ \quad && 
\quad $(3.2 \pm 1.9)\times10^{-1}$ \quad &&
\quad $\Mb^3\vb^5$ \quad &\cr
height10pt && \omit && \omit && \omit &\cr
\tablerule}} $$
\centerline{Table I.  Color-octet matrix elements}
\vfill\eject
$$ \vbox{\offinterlineskip
\def\tablerule{\noalign{\hrule}}
\hrule
\halign {\vrule#& \strut#&
\ \hfil#\hfil & \vrule#&
\ \hfil#\hfil & \vrule#&
\ \hfil#\hfil & \vrule# \cr
\tablerule%
height10pt && \omit && \omit && \omit &\cr
&& Color-Octet Matrix Element && Numerical && NRQCD &\cr
&& Linear Combination && Value $\left({\rm GeV}^3\right)$ 
&&\quad Scaling Order\quad&\cr
height10pt && \omit && \omit && \omit & \cr
\tablerule
height10pt && \omit && \omit && \omit &\cr
&& \quad $\displaystyle{\langle 0|O_8^{J/\psi}({}^3P_0)|0\rangle
\over\Mc^2} + {\langle 0|O_8^{J/\psi}({}^1S_0)|0\rangle\over3}$ \quad
&& \quad $(2.2\pm 0.5)\times 10^{-2}$ \quad
&& \quad $\Mc^3\vc^7$ \quad &\cr
height10pt && \omit && \omit && \omit &\cr
&& \quad $\displaystyle{\langle 0|O_8^{\psi'}({}^3P_0)|0\rangle
\over\Mc^2} + {\langle 0|O_8^{\psi'}({}^1S_0)|0\rangle\over3}$ \quad
&& \quad $(5.9 \pm 1.9)\times 10^{-3}$ \quad
&& \quad $\Mc^3\vc^7$ \quad &\cr
height10pt && \omit && \omit && \omit &\cr
&& \quad $\displaystyle{\langle 0|O_8^{\Upsilon(1S)}({}^3P_0)|
0\rangle\over\Mb^2} + {\langle 0|O_8^{\Upsilon(1S)}({}^1S_0)|0\rangle\over5}$ 
\quad && \quad $(7.9 \pm 10.0)\times 10^{-3}$ \quad
&& \quad $\Mb^3\vb^7$ \quad &\cr
height10pt && \omit && \omit && \omit &\cr
&& \quad $\displaystyle{\langle 0|O_8^{\Upsilon(2S)}({}^3P_0)|
0\rangle\over\Mb^2} + {\langle 0|O_8^{\Upsilon(2S)}({}^1S_0)|0\rangle\over5}$
\quad && \quad $(9.1\pm 7.2)\times 10^{-3}$ \quad
&& \quad $\Mb^3\vb^7$ \quad &\cr
height10pt && \omit && \omit && \omit &\cr
\tablerule}} $$
\centerline{Table II.  Color-octet matrix element linear combinations}
\bigskip\noindent
from the data along with their scaling 
dependence upon the heavy quark mass $\MQ$ and velocity $\vQ$.  The values for 
all the charmonia matrix elements were derived directly from 
the CDF $\Jpsi$ and $\psi'$ data.  On the other hand, insufficient 
experimental information exists to independently extract 
$\langle 0|O_8^{\Upsilon(nS)}({}^3S_1)|0\rangle$ and
$\langle 0|O_8^{\chi_{b1}(nP)}({}^3S_1)|0\rangle$ in the bottomonia sector.  
We therefore determined the latter from the Upsilon data after having scaled 
up the former from the corresponding Psi color-octet matrix elements 
using NRQCD power counting rules.  The remaining
$\langle 0|O_8^{\Upsilon(nS)}({}^3P_0)| 0\rangle / \Mb^2 + 
\langle 0|O_8^{\Upsilon(nS)}({}^1S_0)|0\rangle / 5$ linear combinations were 
obtained directly from the bottomonia cross section data.

	The error bars listed in tables~I and II are statistical and do 
not reflect systematic uncertainties in heavy quark masses, color-singlet 
radial wavefunctions, parton distribution functions and next-to-leading order 
corrections.  The magnitudes of all these different sources of uncertainty can 
be estimated.  For example, the different charm and bottom quark mass values 
which enter into the power law, logarithmic, Coulomb plus linear and QCD 
motivated Buchm\"uller-Tye potentials tabulated in ref.~\Quigg\ span 
the ranges $1.48 \GeV \le \mc \le 1.84 \GeV$ and $4.88 \GeV \le \mb \le 5.18 
\GeV$.  These intervals may be regarded as setting reasonable bounds for 
the heavy quark mass parameters.  The spread in values for radial 
wavefunctions at the origin calculated in these four different potential models 
similarly provides an approximate indication of color-singlet matrix element 
uncertainties.  Systematic errors which arise from parton distribution 
functions and higher order QCD corrections can also be assessed by performing 
several fits with different choices of distribution functions and 
renormalization scale.  We have not attempted to carry out a detailed 
analysis of the combined impact of all these systematic uncertainties.  Our 
color-octet matrix element values therefore represent reasonable estimates 
rather than precise predictions. 

	Comparing the numbers in table~I with their predecessors in table~II 
of ref.~\ChoLeibov, we see that the $\QQbaroctettripS$ matrix element values 
have all diminished.  This is not surprising, for some color-octet 
contributions to quarkonia production are now taken into account by the 
$\QQbaroctettripP$ and $\QQbaroctetsingS$ channels.  We also observe that the 
NRQCD counting rules are more faithfully followed by some matrix elements than 
others.  For instance, the magnitudes of $\langle\CO_8^{\Jpsi}(^3S_1)\rangle$, 
$\langle\CO_8^{\psi'}(^3S_1)\rangle$ and 
$\langle\CO_8^{\psi'}(^3P_0)\rangle/\Mc^2 
+\langle\CO_8^{\psi'}(^1S_0)\rangle/3$ are all mutually consistent with their 
common scaling rule.  On the other hand, 
$\langle\CO_8^{\chi_{c1}}(^3S_1)\rangle$ is somewhat low while
$\langle\CO_8^{\Jpsi}(^3P_0)\rangle/\Mc^2+\langle\CO_8^{\psi'}(^1S_0)\rangle/3$ 
is somewhat high.  Since $\vc^2 \simeq 0.23$ is not very small, none of the 
charmonia NRQCD order-of-magnitude estimates should be overly interpreted.  
We view the general consistency of the fitted matrix elements
with the power counting rules as an encouraging indication that the 
color-octet quarkonia production picture is sound.  

\newsec{Conclusion}

	In this article, we have calculated the cross sections for producing 
colored $L=S=0$ and $L=S=1$ heavy quark-antiquark pairs in hadronic 
collisions.  Intermediate $\QQbaroctetsingS$ and $\QQbaroctettripP$ states 
evolve into $\psiQ$ mesons at the same order in the NRQCD velocity expansion as 
$\QQbaroctettripS$ pairs.  Consistency therefore requires that contributions 
to quarkonia production from all three color-octet channels be considered 
together.  We have found that the full $O(v^7)$ set of color-octet 
distributions yields good fits to prompt Psi and Upsilon data collected at the 
Tevatron.  Numerical values for the long distance matrix elements which can be 
extracted from these data are generally consistent with NRQCD power scaling 
rules.  

	Many of the results in this paper can be applied to a range of other 
interesting problems in quarkonium phenomenology.  In particular, the NRQCD 
matrix elements which we have extracted from CDF data are universal and hold
for color-octet charmonia and bottomonia production at other experimental 
facilities besides the Tevatron.  They can be used, for example, to refine 
the analysis of the $\Jpsi$ differential cross section measured at the CERN 
${\rm S \bar{p}pS}$ collider which was performed in ref.~\CGMP.  It would be 
interesting to see whether disparities between gluon fragmentation predictions 
and UA1 data are diminished by including $\QQbaroctetsingS$ and 
$\QQbaroctettripP$ channels \UAone.  The NRQCD matrix elements can also be 
applied to the study of quarkonia production at lepton colliders.  Gluon 
fragmentation has been shown to represent the largest source of prompt Psi and 
Upsilon vector mesons at LEP \refs{\CKY,\Cho}.  Its incorporation into 
$Z \to \Jpsi$, $Z \to \psi'$ and $Z \to \Upsilon$ branching fractions reduces 
sizable differences between predictions based upon color-singlet heavy quark 
fragmentation and recent LEP measurements.  Color-octet contributions have 
similarly been found to play an important role in charmonia production at CLEO 
\refs{\BraatenChen, \Ko}.  Finally, the color-octet mechanism may eliminate 
disagreements between theory and experiment in fixed target settings.

	We look forward to confronting the color-octet production picture 
with a variety of experimental tests in the near future.  
\foot{The original preprint version of this article contained a section which 
discussed predictions for quarkonia spin alignment resulting from the 
color-octet mechanism.  After some errors in our original polarization 
analysis were pointed out in refs.~\Beneke\ and \BraatenChenII, we removed 
this section.  We will report our revised spin alignment findings in a future 
publication.}
It will be interesting to see how well this simple idea can resolve several 
problems which currently exist in quarkonium physics. 

\bigskip\bigskip\noindent

\bigskip\bigskip\bigskip
\centerline{{\bf Acknowledgments}}
\bigskip

	It is a pleasure to thank Michelangelo Mangano, Vaia Papadimitriou, 
Frank Porter and Mark Wise for helpful discussions. 

\vfill\eject

\appendix{A}{Color-octet squared amplitudes}

	We list below short distance squared amplitudes for $2 \to 1$ and 
$2 \to 2$ scattering processes which mediate color-octet quarkonia production. 
These expressions are averaged over initial spins and colors of the two 
incident partons.  The helicity levels of outgoing $J=1$ and $J=2$ pairs 
are labeled by the subscript $h$.  The total squared amplitudes for creating 
specific quarkonia states are obtained by multiplying these process-independent 
short distance expressions with appropriate long distance NRQCD matrix 
elements. 

\bigskip\noindent
$ q \qbar \to \QQbar[^{2\S +1} L_\J^{(8)}]$ channel:
\eqna\qqbaroctetchannel
$$\eqalignno{
\mathop{{\bar{\sum_{h=0}}}} | \CA(q\qbar \to \QQbaroctettripS) |^2 &= 0
& \qqbaroctetchannel a \cr
\mathop{{\bar{\sum_{|h|=1}}}} | \CA(q\qbar \to \QQbaroctettripS) |^2 &=
{(4 \pi \aS)^2 \over 27 M}  
& \qqbaroctetchannel b \cr} $$

\bigskip\noindent
$ g g  \to \QQbar[^{2\S +1} L_\J^{(8)}]$ channel:
\eqna\ggoctetchannel
$$ \eqalignno{
\mathop{{\bar{\sum}}} | \CA(gg \to \QQbaroctetsingS) |^2 &=
{5 (4\pi \aS)^2 \over 192 M} 
& \ggoctetchannel a \cr
\mathop{{\bar{\sum}}} | \CA(gg \to \QQbaroctetPzero) |^2 &=
{5 (4\pi \aS)^2 \over 16 M^3}
& \ggoctetchannel b \cr
\mathop{{\bar{\sum_{h=0}}}} | \CA(gg \to \QQbaroctetPone) |^2 &= 0 
& \ggoctetchannel c \cr
\mathop{{\bar{\sum_{|h|=1}}}} | \CA(gg \to \QQbaroctetPone) |^2 &= 0
& \ggoctetchannel d \cr
\mathop{{\bar{\sum_{h=0}}}} | \CA(gg \to \QQbaroctetPtwo) |^2 &= 0 
& \ggoctetchannel e \cr
\mathop{{\bar{\sum_{|h|=1}}}} | \CA(gg \to \QQbaroctetPtwo) |^2 &= 0 
& \ggoctetchannel f \cr
\mathop{{\bar{\sum_{|h|=2}}}} | \CA(gg \to \QQbaroctetPtwo) |^2 &=
{(4\pi \aS)^2 \over 12 M^3}
& \ggoctetchannel g \cr} $$

\eject

\noindent
$ q\qbar \to \QQbar[^{2\S +1} L_\J^{(8)}] g$ channel:
\smallskip
\eqna\qqbaroctetgchannel
$$ \eqalignno{
\mathop{{\bar{\sum}}} | \CA(q\qbar \to \QQbaroctetsingS g) |^2 &=
{5 (4\pi \aS)^3 \over 27 M} {\that^2+\uhat^2 \over \shat (\shat-M^2)^2}
& \qqbaroctetgchannel a \cr
\mathop{{\bar{\sum_{h=0}}}} | \CA(q\qbar \to \QQbaroctettripS g) |^2 &=
{8 (4\pi \aS)^3 \over 81 M^3} {M^2 \shat \over (\shat-M^2)^4} 
\bigl[4(\that^2+\uhat^2)-\that\uhat \bigr] 
& \qqbaroctetgchannel b \cr
\mathop{{\bar{\sum_{|h|=1}}}} | \CA(q\qbar \to \QQbaroctettripS g) |^2 &=
{2 (4\pi \aS)^3 \over 81 M^3} {\shat^2+M^4 \over (\shat-M^2)^4} 
{\that^2+\uhat^2 \over \that\uhat} \bigl[4(\that^2+\uhat^2)-\that\uhat \bigr] 
& \qqbaroctetgchannel c \cr 
\mathop{{\bar{\sum}}} | \CA(q\qbar \to \QQbaroctetPzero g) |^2 &=
{20 (4\pi \aS)^3 \over 81 M^3} {(\shat-3 M^2)^2 
(\that^2+\uhat^2) \over \shat(\shat-M^2)^4} 
& \qqbaroctetgchannel d \cr
\mathop{{\bar{\sum_{h=0}}}} | \CA(q\qbar \to \QQbaroctetPone g) |^2 &=
{40 (4\pi \aS)^3 \over 81 M^3} {\shat(\that^2+\uhat^2)
\over (\shat-M^2)^4} 
& \qqbaroctetgchannel e \cr
\mathop{{\bar{\sum_{|h|=1}}}} | \CA(q\qbar \to \QQbaroctetPone g) |^2 &=
{160 (4\pi \aS)^3 \over 81 M^3} {M^2 \that \uhat 
\over (\shat-M^2)^4} 
& \qqbaroctetgchannel f \cr
\mathop{{\bar{\sum_{h=0}}}} | \CA(q\qbar \to \QQbaroctetPtwo g) |^2 &=
{8 (4\pi \aS)^3 \over 81 M^3} {\shat(\that^2+\uhat^2)
\over (\shat-M^2)^4} 
& \qqbaroctetgchannel g \cr
\mathop{{\bar{\sum_{|h|=1}}}} | \CA(q\qbar \to \QQbaroctetPtwo g) |^2 &=
{32 (4\pi \aS)^3 \over 27 M^3} {M^2 \that\uhat 
\over (\shat-M^2)^4} 
& \qqbaroctetgchannel h \cr
\mathop{{\bar{\sum_{|h|=2}}}} | \CA(q\qbar \to \QQbaroctetPtwo g) |^2 &=
{16 (4\pi \aS)^3 \over 27 M^3} {M^4 (\that^2+\uhat^2)
\over \shat(\shat-M^2)^4}
& \qqbaroctetgchannel i \cr }$$

\vfill\eject

\noindent
$ g q \to \QQbar[^{2\S+1} L_\J^{(8)}] q$ channel:
\medskip
\eqna\gqoctetqchannel
$$ \eqalignno{
\mathop{{\bar{\sum}}} | \CA(gq \to \QQbaroctetsingS q) |^2 &=
-{5 (4\pi \aS)^3 \over 72 M} {\shat^2+\uhat^2 \over \that (\that-M^2)^2}
& \gqoctetqchannel a \cr
\mathop{{\bar{\sum_{h=0}}}} | \CA(gq \to \QQbaroctettripS q) |^2 &=
-{(4\pi \aS)^3 \over 54 M^3} {M^2 \that 
\bigl[4(\shat^2+\uhat^2)-\shat\uhat\bigr] \over 
\bigl[(\shat-M^2)(\that-M^2)\bigr]^2}  
& \gqoctetqchannel b \cr
\mathop{{\bar{\sum_{|h|=1}}}} | \CA(gq \to \QQbaroctettripS q) |^2 &=
-{(4\pi \aS)^3 \over 108 M^3} {(\shat^2+\uhat^2+2 M^2 \that)(\shat-M^2)^2 
- 2 M^2 \shat\that\uhat \over \shat\uhat 
\bigl[(\shat-M^2)(\that-M^2)\bigr]^2} & \cr
& \quad\qquad \times \bigl[4(\shat^2+\uhat^2)-\shat\uhat\bigr] 
& \gqoctetqchannel c \cr
\mathop{{\bar{\sum}}} | \CA(gq \to \QQbaroctetPzero q) |^2 &=
-{5 (4\pi \aS)^3 \over 54 M^3} {(\that-3 M^2)^2 
(\shat^2+\uhat^2) \over \that(\that-M^2)^4} 
& \gqoctetqchannel d \cr
\mathop{{\bar{\sum_{h=0}}}} | \CA(gq \to \QQbaroctetPone q) |^2 &=
-{5 (4\pi \aS)^3 \over 27 M^3} {\that \bigl[
\shat^2(\shat-M^2)^2 + \uhat^2(\shat+M^2)^2 \bigr] \over
(\that-M^2)^4 (\shat-M^2)^2} 
& \gqoctetqchannel e \cr
\mathop{{\bar{\sum_{|h|=1}}}} | \CA(gq \to \QQbaroctetPone q) |^2 &=
-{20 (4\pi \aS)^3 \over 27 M^3} {M^2 \shat \uhat 
(\that^2+\that\uhat+\uhat^2) \over (\that-M^2)^4 (\shat-M^2)^2} 
& \gqoctetqchannel f \cr
\mathop{{\bar{\sum_{h=0}}}} | \CA(gq \to \QQbaroctetPtwo q) |^2 &=
-{(4\pi \aS)^3 \over 27 M^3} {\that \over (\that-M^2)^4}
\cr
& \quad\qquad \times \bigl[ \shat^2+\uhat^2+12 M^2 \shat \uhat^2 
{\shat^2+M^2 \shat + M^4 \over
(\shat-M^2)^4} \bigr] 
& \gqoctetqchannel g \cr
\mathop{{\bar{\sum_{|h|=1}}}} | \CA(gq \to \QQbaroctetPtwo q) |^2 &=
-{4 (4\pi \aS)^3 \over 9 M^3} {M^2 \shat\uhat 
\over (\that-M^2)^4} \cr
& \quad\qquad \times {(\shat-M^2)^2 (\shat^2+M^4) - (\shat+M^2)^2 \that\uhat 
\over (\shat-M^2)^4}
& \gqoctetqchannel h \cr
\mathop{{\bar{\sum_{|h|=2}}}} | \CA(gq \to \QQbaroctetPtwo q) |^2 &=
-{2 (4\pi \aS)^3 \over 9 M^3} {M^4 
\over \that(\that-M^2)^4} \cr
& \quad\qquad \times \bigl[ \shat^2+\uhat^2 + 2 \shat^2 \that \uhat
{(\shat-M^2)(2 \that+\uhat) - \uhat^2 \over (\shat-M^2)^4} \bigr]  
& \gqoctetqchannel i \cr } $$

\vfill\eject

\noindent
$ g g \to \QQbar[^{2\S+1} L_\J^{(8)}] g$ channel:
\foot{The $gg \to \QQbaroctettripP\ g$ squared amplitudes are expressed in 
terms of the variables $\shat$ and $\zhat \equiv \sqrt{\that\uhat}$.}

\medskip
\eqna\ggoctetgchannel
$$ \eqalignno{
& \mathop{{\bar{\sum}}} | \CA(gg \to \QQbaroctetsingS g) |^2 =
{5 (4\pi \aS)^3 \over 16 M} \bigl[ \shat^2 (\shat-M^2)^2 + \shat \that\uhat
(M^2-2\shat) + (\that\uhat)^2 \bigr] \cr
& \qquad\qquad \times { (\shat^2-M^2 \shat + M^4)^2
-\that\uhat (2 \that^2 + 3 \that\uhat + 2 \uhat^2) 
\over \shat\that\uhat \bigl[ (\shat-M^2) (\that-M^2) (\uhat-M^2) \bigr]^2} 
& \ggoctetgchannel a \cr
& & \cr
& & \cr
& \mathop{{\bar{\sum_{h=0}}}} | \CA(g g \to \QQbaroctettripS g) |^2 =
-{(4\pi \aS)^3 \over 144 M^3}
{2 M^2 \shat \over (\shat-M^2)^2} (\that^2+\uhat^2) \that\uhat & \cr
& \qquad \qquad \times {27(\shat\that+\that\uhat+\uhat\shat) 
-19M^4 \over \bigl[(\shat-M^2)(\that-M^2)(\uhat-M^2)\bigr]^2} 
& \ggoctetgchannel b \cr
& & \cr
& & \cr
& \mathop{{\bar{\sum_{|h|=1}}}} | \CA(g g \to \QQbaroctettripS g) |^2 =
-{(4\pi \aS)^3 \over 144 M^3} {\shat^2 \over (\shat-M^2)^2} 
\bigl[(\shat-M^2)^4+\that^4+\uhat^4+2 M^4 \bigl({\that\uhat \over \shat} 
\bigr)^2 \bigr] & \cr
& \qquad \qquad \times {27(\shat\that+\that\uhat+\uhat\shat) 
-19M^4 \over \bigl[(\shat-M^2)(\that-M^2)(\uhat-M^2)\bigr]^2} 
& \ggoctetgchannel c \cr
& & \cr
& & \cr
& \mathop{{\bar{\sum}}} | \CA(gg \to \QQbaroctetPzero g) |^2 =
{5 (4\pi \aS)^3 \over 12 M^3} \cr
& \qquad\qquad \times \Bigl\{\shat^2 \zhat^4 (\shat^2-\zhat^2)^4 
+ M^2 \shat \zhat^2 (\shat^2-\zhat^2)^2 (3\shat^2-2\zhat^2)
(2\shat^4 - 6 \shat^2 \zhat^2 + 3 \zhat^4) \cr
& \qquad\qquad\qquad + M^4 \bigl[ 9\shat^{12} - 84 \shat^{10} \zhat^2 
  + 265 \shat^8 \zhat^4 
  - 382 \shat^6 \zhat^6 + 276 \shat^4 \zhat^8 - 88 \shat^2 \zhat^{10} 
  + 9 \zhat^{12} \bigr] 
\cr 
& \qquad\qquad\qquad - M^6 \shat \bigl[ 54 \shat^{10} - 357 \shat^8 \zhat^2 
 + 844 \shat^6 \zhat^4 - 898 \shat^4 \zhat^6 + 439 \shat^2 \zhat^8 
 - 81 \zhat^{10} \bigr] \cr
& \qquad\qquad\qquad + M^8 \bigl[ 153 \shat^{10} - 798 \shat^8 \zhat^2 
  + 1415 \shat^6 \zhat^4
  - 1041 \shat^4 \zhat^6 + 301 \shat^2 \zhat^8 - 18 \zhat^{10} \bigr] \cr
& \qquad\qquad\qquad -M^{10} \shat \bigl[ 270 \shat^8 - 1089 \shat^6 \zhat^2 
  + 1365 \shat^4 \zhat^4 - 616 \shat^2 \zhat^6 + 87 \zhat^8 \bigr] \cr
& \qquad\qquad\qquad + M^{12} \bigl[ 324 \shat^8 - 951 \shat^6 \zhat^2 
  + 769 \shat^4 \zhat^4 
  - 189 \shat^2 \zhat^6 + 9 \zhat^8 \bigr] \cr
& \qquad\qquad\qquad - 9 M^{14} \shat (6 \shat^2 - \zhat^2) (5 \shat^4 
  - 9 \shat^2 \zhat^2 + 3 \zhat^4) \cr
& \qquad\qquad\qquad + 3 M^{16} \shat^2 (51 \shat^4-59 \shat^2 \zhat^2 
  + 12 \zhat^4)
  - 27 M^{18} \shat^3 (2\shat^2 - \zhat^2) + 9 M^{20} \shat^4 \Bigr\} \cr
 & \qquad\qquad / \bigl[\shat \zhat^2 (\shat-M^2)^4 (\shat M^2 + \zhat^2)^4 
 \bigr]  
& \ggoctetgchannel d \cr}$$
\vfill\eject

$$ \eqalignno{
& \mathop{{\bar{\sum_{h=0}}}} | \CA(gg \to \QQbaroctetPone g) |^2 =
{5 (4\pi \aS)^3 \over 6 M^3} \cr
& \qquad\qquad \times \shat \zhat^2 \bigl[(\shat^2-\zhat^2)^2 
  - 2 M^2 \shat \zhat^2
  - M^4 (\shat^2+2 \zhat^2) + M^8 \bigr] \cr
& \qquad\qquad \times \bigl[(\shat^2-\zhat^2)^2 - M^2 \shat (2\shat^2 - 
  \zhat^2) + M^4 \shat^2 \bigr] / \bigl[(\shat-M^2)^4 (\shat M^2 + \zhat^2)^4 
  \bigr]
& \ggoctetgchannel e \cr
& & \cr
& \mathop{{\bar{\sum_{|h|=1}}}} | \CA(gg \to \QQbaroctetPone g) |^2 =
{5 (4\pi \aS)^3 \over 6 M^3} \cr
& \qquad\qquad \times M^2 \Bigl\{ 2(\shat^2-\zhat^2)^2 (\shat^6-4 \shat^4 
  \zhat^2 + \shat^2 \zhat^4 - \zhat^6) \cr
& \qquad\qquad\qquad - M^2 \shat (2 \shat^2-\zhat^2) (5 \shat^6 - 17 \shat^4 
  \zhat^2
  + 9 \shat^2 \zhat^4 - \zhat^6 ) \cr
& \qquad\qquad\qquad  + M^4 ( 21\shat^8 - 49 \shat^6 \zhat^2 
  + 21 \shat^4 \zhat^4
  - 4 \shat^2 \zhat^6 + \zhat^8) \cr
& \qquad\qquad\qquad - M^6 \shat (24 \shat^6 - 30 \shat^4 \zhat^2 
  + 6 \shat^2 \zhat^4 - \zhat^6) \cr
& \qquad\qquad\qquad + M^8 \shat^2 (16 \shat^4 - 9 \shat^2 \zhat^2 + 2 \zhat^4)
  - M^{10} \shat^3 (6 \shat^2 - \zhat^2) + M^{12} \shat^4 \Bigr\} \cr
& \qquad\qquad / \bigl[(\shat-M^2)^4 (\shat M^2 + \zhat^2)^4 \bigr]
& \ggoctetgchannel f \cr
& & \cr
& \mathop{{\bar{\sum_{h=0}}}} | \CA(gg \to \QQbaroctetPtwo g) |^2 =
{(4\pi \aS)^3 \over 6 M^3} \cr
& \qquad\qquad \times \shat \zhat^2 \Bigl\{
\shat^2 (\shat^2-\zhat^2)^4 - M^2 \shat \zhat^2 (\shat^2-\zhat^2)^2 
  (11 \shat^2+2 \zhat^2) \cr
& \qquad\qquad\qquad + M^4 \bigl[ \shat^8 - 12 \shat^6 \zhat^2 
  + 41 \shat^4 \zhat^4
  - 20 \shat^2 \zhat^6 + \zhat^8 \bigr] \cr
& \qquad\qquad\qquad - M^6 \shat \bigl[ 4\shat^6 - 26 \shat^4 \zhat^2 
  - \shat^2 \zhat^4 - 5 \zhat^6 \bigr] \cr
& \qquad\qquad\qquad + M^8 \bigl[ 29 \shat^6 - 114 \shat^4 \zhat^2 
  + 108 \shat^2
  \zhat^4 - 10 \zhat^6 \bigr] \cr
& \qquad\qquad\qquad - M^{10} \shat \bigl[ 65 \shat^4 - 104 \shat^2 \zhat^2
  -33 \zhat^4 \bigr] + M^{12} \bigl[ 54 \shat^4 - 20 \shat^2 \zhat^2 
  + 7 \zhat^4 \bigr] \cr
& \qquad\qquad\qquad - M^{14} \shat (23 \shat^2 + 5 \zhat^2) 
  + 7 M^{16} \shat^2 \Bigr\} 
 / \bigl[(\shat-M^2)^6 (\shat M^2 + \zhat^2)^4 \bigr]
& \ggoctetgchannel g \cr}$$
\vfill\eject

$$ \eqalignno{
& \mathop{{\bar{\sum_{|h|=1}}}} | \CA(gg \to \QQbaroctetPtwo g) |^2 =
{(4\pi \aS)^3 \over 2 M^3} \cr
& \qquad\qquad \times M^2 \Bigl\{
2 \shat^2 (\shat^2-\zhat^2)^2 (\shat^6 - 4 \shat^4 \zhat^2 + \shat^2 \zhat^4 
  - \zhat^6) \cr
& \qquad\qquad\qquad - M^2 \shat \bigl[ 10 \shat^{10} - 37 \shat^8 \zhat^2
  + 19 \shat^6 \zhat^4 + 11 \shat^4 \zhat^6 - \shat^2 \zhat^8 
  - 4 \zhat^{10} \bigr] \cr
& \qquad\qquad\qquad + M^4 \bigl[ 25 \shat^{10} - 61 \shat^8 \zhat^2 
  + 27 \shat^6
  \zhat^4 - 34 \shat^4 \zhat^6 + 23 \shat^2 \zhat^8 - 2 \zhat^{10} \bigr] \cr
& \qquad\qquad\qquad - M^6 \shat \bigl[ 42 \shat^8 - 77 \shat^6 \zhat^2 + 
  41 \shat^4 \zhat^4 - 22 \shat^2 \zhat^6 + 17 \zhat^8 \bigr] \cr
& \qquad\qquad\qquad + M^8 \bigl[ 53 \shat^8 - 88 \shat^6 \zhat^2 
  + 69 \shat^4 \zhat^4
  - 68 \shat^2 \zhat^6 + 3 \zhat^8 \bigr] \cr
& \qquad\qquad\qquad - M^{10} \shat \bigl[ 54 \shat^6 - 85 \shat^4 \zhat^2 +
  60 \shat^2 \zhat^4 - 9 \zhat^6 \bigr] + M^{12} \shat^2 \bigl[ 43 \shat^4 - 
  47 \shat^2 \zhat^2 + 20 \zhat^4 \bigr] \cr
& \qquad\qquad\qquad - M^{14} \shat^3 (22 \shat^2 - 9 \zhat^2) 
  + 5 M^{16} \shat^4 \Bigr\} 
 / \bigl[(\shat-M^2)^6 (\shat M^2+\zhat^2)^4 \bigr]
& \ggoctetgchannel h \cr
& & \cr
& \mathop{{\bar{\sum_{|h|=2}}}} | \CA(gg \to \QQbaroctetPtwo g) |^2 =
{(4\pi \aS)^3 \over 2 M^3} \cr
& \qquad\qquad \times M^4 \Bigl\{
2 \shat^2 \bigl[ \shat^{12} - 8 \shat^{10} \zhat^2 + 22 \shat^8 \zhat^4 
  - 24 \shat^6
  \zhat^6 + 10 \shat^4 \zhat^8 - 3 \shat^2 \zhat^{10} + \zhat^{12} \bigr] \cr
& \qquad\qquad\qquad - M^2 \shat \bigl[ 16 \shat^{12} - 102 \shat^{10} \zhat^2
  + 210 \shat^8 \zhat^4 - 153 \shat^6 \zhat^6 + 36 \shat^4 \zhat^8 
  - 6 \shat^2 \zhat^{10}
  + 4 \zhat^{12} \bigr] \cr
& \qquad\qquad\qquad + M^4 \bigl[ 60 \shat^{12} - 306 \shat^{10} \zhat^2
  + 482 \shat^8 \zhat^4 - 271 \shat^6 \zhat^6 + 77 \shat^4 \zhat^8 
  - 18 \shat^2 \zhat^{10}
  + 2 \zhat^{12} \bigr] \cr
& \qquad\qquad\qquad - M^6 \shat \bigl[ 140 \shat^{10} - 573 \shat^8 \zhat^2
  + 710 \shat^6 \zhat^4 - 344 \shat^4 \zhat^6 + 91 \shat^2 \zhat^8 
  - 18 \zhat^{10} \bigr] \cr
& \qquad\qquad\qquad + M^8 \bigl[ 226 \shat^{10} - 741 \shat^8 \zhat^2 + 737 
  \shat^6 \zhat^4 - 310 \shat^4 \zhat^6 + 77 \shat^2 \zhat^8 
  - 4 \zhat^{10} \bigr] \cr
& \qquad\qquad\qquad - M^{10} \shat \bigl[ 264 \shat^8 - 686 \shat^6 \zhat^2
  + 541 \shat^4 \zhat^4 - 177 \shat^2 \zhat^6 + 25 \zhat^8 \bigr] \cr
& \qquad\qquad\qquad + M^{12} \bigl[ 226 \shat^8 - 452 \shat^6 \zhat^2 + 
  261 \shat^4 \zhat^4 - 55 \shat^2 \zhat^6 + 2 \zhat^8 \bigr] \cr
& \qquad\qquad\qquad - M^{14} \shat \bigl[ 140 \shat^6 - 201 \shat^4 \zhat^2
  + 71 \shat^2 \zhat^4 - 6 \zhat^6 \bigr] \cr
& \qquad\qquad\qquad + M^{16} \shat^2 \bigl[ 60 \shat^4 - 53 \shat^2 \zhat^2
  + 8 \zhat^4 \bigr] - 2 M^{18} \shat^3 \bigl[ 8 \shat^2 - 3 \zhat^2 \bigr]
  + 2 M^{20} \shat^4 \Bigr\} \cr
& \qquad\qquad / \bigl[ \shat \zhat^2 (\shat-M^2)^6 (\shat M^2 + \zhat^2)^4 
  \bigr] 
& \ggoctetgchannel i \cr
} $$

\listfigs
\listrefs
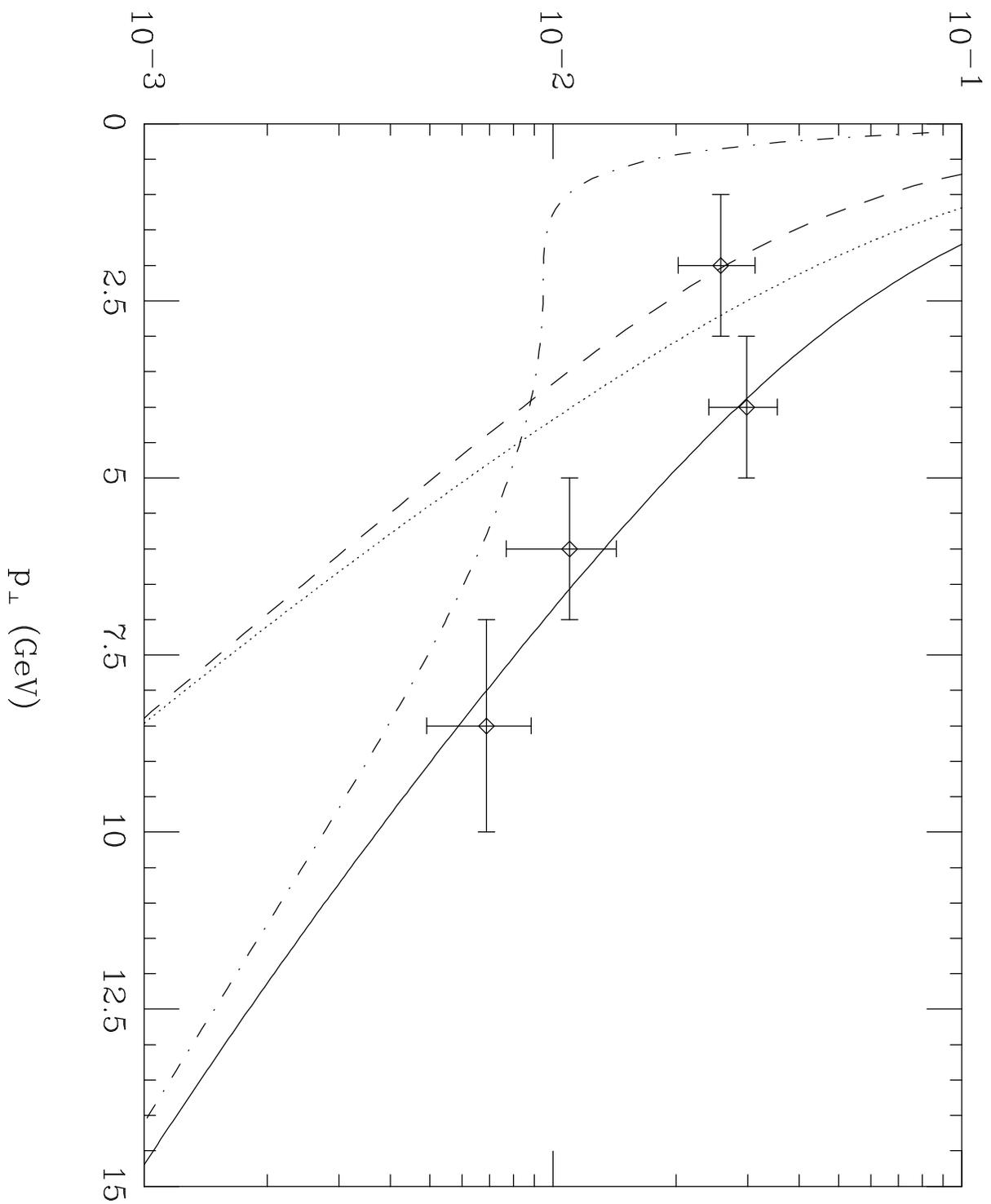
\bye